\documentclass[12pt]{article}

\usepackage{epsf}

\usepackage{rotate}

\renewcommand{\thefootnote}{\fnsymbol{footnote}}

\begin{document}


\begin{titlepage}
\begin{flushright}
\begin{tabular}{l}
DESY 00--051\\
TUM--HEP--367/00\\
hep--ph/0003323\\
March 2000
\end{tabular}
\end{flushright}

\vspace*{1.8truecm}

\begin{center}
\boldmath
{\Large \bf Constraints on the CKM Angle $\gamma$ and}

\vspace*{0.3truecm}

{\Large \bf Strong Phases from $B\to\pi K$ Decays}
\unboldmath

\vspace*{2.5cm}

{\sc{\large Andrzej J. Buras}}${}^{1,}$\footnote{E-mail: 
{\tt Andrzej.Buras@feynman.t30.physik.tu-muenchen.de}} \,\, {\large and }\,\,
{\sc{\large Robert Fleischer}}${}^{2,}$\footnote{E-mail: 
{\tt Robert.Fleischer@desy.de}}\\[0.8cm]
\vspace*{0.1cm} ${}^1${\it Technische Universit\"at M\"unchen, 
Physik Department, D--85748 Garching, Germany}\\[0.3cm]
\vspace*{0.1cm} ${}^2${\it Deutsches Elektronen-Synchrotron DESY, 
Notkestr.\ 85, D--22607 Hamburg, Germany}

\vspace{2.4truecm}

{\large\bf Abstract\\[10pt]} \parbox[t]{\textwidth}{
As we pointed out recently, the neutral decays $B_d\to\pi^\mp K^\pm$ and 
$B_d\to\pi^0K$ may provide non-trivial bounds on the CKM angle $\gamma$. 
In this paper, we reconsider this approach in the light of recent CLEO data, 
which look very interesting. In particular, the results for the corresponding 
CP-averaged branching ratios are in favour of strong constraints on $\gamma$, 
where the second quadrant is preferred. Such a situation would be in conflict 
with the standard analysis of the unitarity triangle. Moreover, constraints 
on a CP-conserving strong phase $\delta_{\rm n}$ are in favour of a negative
value of $\cos\delta_{\rm n}$, which would be in conflict with the 
factorization expectation. In addition, there seems to be an interesting 
discrepancy with the bounds that are implied by the charged $B\to\pi K$ 
system: whereas these decays favour a range for $\gamma$ that is similar 
to that of the neutral modes, they point towards a positive value of
$\cos\delta_{\rm c}$, which would be in conflict with the expectation of 
equal signs for $\cos\delta_{\rm n}$ and $\cos\delta_{\rm c}$. If future 
data should confirm this ``puzzle'', it may be an indication for new-physics 
contributions to the electroweak penguin sector, or a manifestation of 
large non-factorizable $SU(3)$-breaking effects.
}

\vskip1.5cm

\end{center}

\end{titlepage}

\thispagestyle{empty}
\vbox{}
\newpage
 
\setcounter{page}{1}

\setcounter{footnote}{0}
\renewcommand{\thefootnote}{\arabic{footnote}}

\section{Introduction}\label{intro}
In order to probe the angle $\gamma$ of the unitarity triangle of the 
Cabibbo--Kobayashi--Maskawa (CKM) matrix at the $B$-factories, 
$B\to\pi K$ decays play an outstanding role. Remarkably, already
CP-averaged branching ratios of such channels may imply very non-trivial
constraints on $\gamma$. So far, the studies of these bounds have focussed 
on the following two systems: $B_d\to\pi^\mp K^\pm$, $B^\pm\to\pi^\pm K$ 
\cite{FM}, and $B^\pm\to\pi^0K^\pm$, $B^\pm\to\pi^\pm K$ \cite{NR}; they
have received a lot of attention in the literature. In a recent paper 
\cite{BF}, we pointed out that also the neutral decays $B_d\to\pi^\mp K^\pm$ 
and $B_d\to\pi^0 K$ may be interesting in this respect, and presented 
a general formalism, allowing us to describe all three $B\to\pi K$ systems 
within the same theoretical framework. Since the CLEO collaboration has
reported the observation of the $B_d\to\pi^0 K$ channel in the summer of
1999, which finalizes the search for all four $B\to\pi K$ modes, we have 
reanalysed our approach in view of these new data. It turns out that 
the new CLEO results~\cite{CLEO1} favour interesting bounds on 
$\gamma$ from the neutral $B\to\pi K$ decays. Here the key quantities 
are the following ratios of CP-averaged branching ratios \cite{BF}:
\begin{eqnarray}
R&\equiv&\frac{\mbox{BR}(B^0_d\to\pi^-K^+)+
\mbox{BR}(\overline{B^0_d}\to\pi^+K^-)}{\mbox{BR}(B^+\to\pi^+K^0)+
\mbox{BR}(B^-\to\pi^-\overline{K^0})}=0.95\pm0.28\label{R-def}\\
R_{\rm c}&\equiv&2\left[\frac{\mbox{BR}(B^+\to\pi^0K^+)+
\mbox{BR}(B^-\to\pi^0K^-)}{\mbox{BR}(B^+\to\pi^+K^0)+
\mbox{BR}(B^-\to\pi^-\overline{K^0})}\right]=1.27\pm0.47\label{Rc-def}\\
R_{\rm n}&\equiv&\frac{1}{2}\left[\frac{\mbox{BR}(B^0_d\to\pi^-K^+)+
\mbox{BR}(\overline{B^0_d}\to\pi^+K^-)}{\mbox{BR}(B^0_d\to\pi^0K^0)+
\mbox{BR}(\overline{B^0_d}\to\pi^0\overline{K^0})}\right]=
0.59\pm0.27,\label{Rn-def}
\end{eqnarray}
where the factors of 2 and 1/2 have been introduced to absorb the $\sqrt{2}$
factors originating from the wavefunctions of the neutral pions; the 
errors of the experimental results given in Ref.~\cite{CLEO1} have been
added in quadrature. If these ratios are found to be {\it smaller} than one, 
they can be converted directly into constraints on $\gamma$ without any 
additional information. When the $B_d\to\pi^\mp K^\pm$, $B^\pm\to\pi^\pm K$ 
channels were observed in 1997 by the CLEO collaboration, the first results 
gave $R=0.65\pm0.40$, and the bound on $\gamma$ presented in Ref.~\cite{FM} 
led to great excitement in the $B$-physics community. In the case of
$R_{\rm n}$, we now face a similarly exciting possibility, which we will
discuss in more detail in this paper. However, in comparison with the 
original bound derived in \cite{FM}, the neutral strategy has certain 
theoretical advantages, connected mainly with the impact of rescattering 
processes \cite{FSI}--\cite{FKNP} and electroweak penguin topologies. 

If one of the ratios $R_{({\rm c,n})}$ specified in 
(\ref{R-def})--(\ref{Rn-def}) is found to be larger than one, additional 
experimental information is required to constrain $\gamma$. To this end, 
we have then to fix -- sloppily speaking -- certain ratios of ``tree'' 
to ``penguin'' amplitudes. Such an input allows us also to obtain stronger 
constraints on $\gamma$ in the case of $R_{({\rm c,n})}<1$. The least
fortunate case for the bounds on $\gamma$ would be $R_{({\rm c,n})}$ close
to 1. If CP-violating asymmetries in the channels appearing in the numerators 
in (\ref{R-def})--(\ref{Rn-def}) can be measured, it is possible to go beyond 
the bounds on $\gamma$ and to determine this angle, also in the case
of $R_{({\rm c,n})}=1$. A first analysis of such CP asymmetries has recently 
been performed by the CLEO collaboration \cite{CLEO2}, where all results 
are unfortunately still consistent with zero. It is also possible to obtain 
theoretical upper bounds on such CP asymmetries. For instance, the ratio of 
the measured CP-averaged $B_d\to\pi^+\pi^-$ and $B_d\to\pi^\mp K^\pm$ 
branching ratios implies $|{\cal A}_{\rm CP}^{\rm dir}(B_d\to\pi^\mp K^\pm)|
\mathrel{\hbox{\rlap{\hbox{\lower4pt\hbox{$\sim$}}}\hbox{$<$}}}0.3$ 
\cite{RF-Bpipi}.

It is an interesting feature of the bounds on $\gamma$ that they prefer
values in the second quadrant, which would be in conflict with the standard 
analysis of the unitarity triangle \cite{UT-fits}. Other arguments for 
$\cos\gamma<0$ using $B\to PP$, $PV$ and $VV$ decays were recently given 
in \cite{HY} (see also \cite{RF-Bpipi}). We would like to point out that, 
in addition to the bounds on $\gamma$, one may also derive constraints on 
CP-conserving strong phases $\delta_{\rm n}$ and $\delta_{\rm c}$ from the 
neutral and charged $B\to\pi K$ decays, respectively. Whereas the present 
CLEO data favour a positive value of $\cos\delta_{\rm c}$, as is expected 
in the factorization approximation, they point towards a negative value of 
$\cos\delta_{\rm n}$. However, on the basis of simple dynamical 
considerations, one would expect that $\delta_{\rm n}$ and $\delta_{\rm c}$ 
do not differ dramatically from each other. The present data do of course 
not allow us to draw any definite conclusions. However, if the future data 
should confirm this interesting ``puzzle'', it may be an indication for 
new-physics contributions to the electroweak penguin sector, or a 
manifestation of large non-factorizable $SU(3)$-breaking effects.

The outline of this paper is as follows: in Section~\ref{sec:strat},
we repeat briefly the general formalism developed in \cite{BF}. The bounds
on $\gamma$ are discussed in view of the recent CLEO data in 
Section~\ref{sec:bounds}, where we also have a brief look at constraints 
in the $\overline{\varrho}$--$\overline{\eta}$ plane of the Wolfenstein 
parameters \cite{wolf}, generalized as in Ref.~\cite{BLO}. In 
Section~\ref{sec:delta}, we turn to the constraints on the strong
phases $\delta_{\rm n}$ and $\delta_{\rm c}$. Finally, a few concluding 
remarks are given in Section~\ref{sec:concl}.

\section{General Formalism}\label{sec:strat}
The starting point of our description of the neutral $B\to\pi K$ system
is the following isospin relation:
\begin{equation}\label{iso1}
\sqrt{2}\,A(B^0_d\to\pi^0K^0)\,+\,A(B^0_d\to\pi^-K^+)=
-\left[(T+C)\,+\,P_{\rm ew}\right]\equiv 3\,A_{3/2},
\end{equation}
where the combination $(T+C)$ originates from colour-allowed and 
colour-suppressed $\bar b\to\bar uu\bar s$ tree-diagram-like topologies, 
$P_{\rm ew}$ is due to electroweak penguin constributions, and $A_{3/2}$
reminds us that there is only an $I=3/2$ isospin component present in 
(\ref{iso1}). Within the Standard Model, these amplitudes can be 
parametrized as follows:
\begin{equation}\label{Ampl-def}
T+C=|T+C|\,e^{i\delta_{T+C}}\,e^{i\gamma},\quad
P_{\rm ew}=-\,|P_{\rm ew}|e^{i\delta_{\rm ew}},
\end{equation}
where $\delta_{T+C}$ and $\delta_{\rm ew}$ denote CP-conserving strong
phases. For the following considerations, we have to parametrize the 
$B^0_d\to\pi^0K^0$ decay amplitude in an appropriate way. If we make
use of the unitarity of the CKM matrix and employ the Wolfenstein 
parametrization \cite{wolf}, generalized to include
non-leading terms in $\lambda\equiv|V_{us}|=0.22$ \cite{BLO}, we
obtain
\begin{equation}\label{Bnampl}
\sqrt{2}\,A(B^0_d\to\pi^0K^0)\equiv P_{\rm n}=
-\left(1-\frac{\lambda^2}{2}\right)\lambda^2A
\left[1+\rho_{\rm n}\,e^{i\theta_{\rm n}}e^{i\gamma}\right]
{\cal P}_{tc}^{\rm n}\,,
\end{equation}
where $\rho_{\rm n}\,e^{i\theta_{\rm n}}$ takes the form 
\begin{equation}\label{rho-n-def}
\rho_{\rm n}\,e^{i\theta_{\rm n}}=\frac{\lambda^2R_b}{1-\lambda^2}
\left[1-\left(\frac{{\cal P}_{uc}^{\rm n}-{\cal C}}{{\cal P}_{tc}^{\rm n}}
\right)\right].
\end{equation}
Here ${\cal P}_{tc}^{\rm n}\equiv|{\cal P}_{tc}^{\rm n}|\,
e^{i\delta_{tc}^{\rm n}}$ and ${\cal P}_{uc}^{\rm n}$ correspond to 
differences of penguin topologies with internal top and charm and up and 
charm quarks, respectively. The amplitude ${\cal C}$ is due to insertions 
of current--current operators into colour-suppressed tree-diagram-like 
topologies, and 
\begin{equation}\label{Rb-def}
A\equiv\frac{1}{\lambda^2}|V_{cb}|=0.81\pm0.06,\quad
R_b\equiv\frac{1}{\lambda}\left(1-\frac{\lambda^2}{2}\right)
\left|\frac{V_{ub}}{V_{cb}}\right|=\sqrt{\overline{\varrho}^2+
\overline{\eta}^2}=0.41\pm0.07
\end{equation}
are the usual CKM factors. In order to parametrize the observable $R_{\rm n}$ 
defined in (\ref{Rn-def}), it is useful to introduce the following quantities:
\begin{equation}\label{rn-def}
r_{\rm n}\equiv\frac{|T+C|}{\sqrt{\langle|P_{\rm n}|^2\rangle}}\,, 
\quad\delta_{\rm n}\equiv\delta_{T+C}-\delta_{tc}^{\rm n}\,,
\end{equation}
where 
\begin{equation}
\langle|P_{\rm n}|^2\rangle\equiv\frac{1}{2}\left(|P_{\rm n}|^2+
|\overline{P_{\rm n}}|^2\right)
\end{equation} 
is the CP-average of the $B_d^0\to\pi^0K^0$ decay amplitude specified 
in (\ref{Bnampl}). Then we obtain \cite{BF,defan}:
\begin{equation}\label{Rn-exp}
R_{\rm n}=1-\frac{2\,r_{\rm n}}{u_{\rm n}}\left(h_{\rm n}\cos\delta_{\rm n}+
k_{\rm n}\sin\delta_{\rm n}\right)+v^2r_{\rm n}^2,
\end{equation}
where
\begin{eqnarray}
h_{\rm n}&=&\cos\gamma+\rho_{\rm n}\cos\theta_{\rm n}-q\left[
\,\cos\omega+\rho_{\rm n}\cos(\theta_{\rm n}-\omega)\cos\gamma\,
\right]\label{h-def}\\
k_{\rm n}&=&\rho_{\rm n}\sin\theta_{\rm n}+q\left[\,\sin\omega-\rho_{\rm n}
\sin(\theta_{\rm n}-\omega)\cos\gamma\,\right],
\end{eqnarray}
and
\begin{eqnarray}
u_{\rm n}&=&\sqrt{1+2\,\rho_{\rm n}\,\cos\theta_{\rm n}\cos\gamma+
\rho_{\rm n}^2}\\
v&=&\sqrt{1-2\,q\cos\omega\cos\gamma+q^2}\,.\label{v-def}
\end{eqnarray}
Moreover, we have introduced the electroweak penguin parameter
\begin{equation}
q\,e^{i\omega}\equiv\left|\frac{P_{\rm ew}}{T+C}\right|\,
e^{i(\delta_{\rm ew}-\delta_{T+C})},
\end{equation}
which can be fixed theoretically \cite{NR} (see also \cite{PAPIII}).
This interesting observation was made by Neubert and Rosner in the context
of the charged $B\to\pi K$ system. However, as (\ref{iso1}) is also 
satisfied by the corresponding charged combination, the same feature can 
be used in the neutral strategy as well \cite{BF}. To this end, two 
electroweak penguin operators with tiny Wilson coefficients are neglected, 
as well as electroweak penguins with internal up and charm quarks. 
Furthermore, appropriate Fierz transformations of the remaining electroweak 
penguin operators are performed, and the $SU(3)$ flavour symmetry of strong 
interactions is applied. Finally, one arrives at the following 
result \cite{NR}:
\begin{equation}\label{q-expr}
q\,e^{i\omega}=0.63\times\left[\frac{0.41}{R_b}\right],
\end{equation}
where also factorizable $SU(3)$-breaking corrections have been taken into
account. The amplitude $T+C$, i.e.\ the parameter $r_{\rm n}$, 
can be determined with the help of the decay $B^+\to\pi^+\pi^0$ by 
using the $SU(3)$ flavour symmetry of strong interactions~\cite{GRL}:
\begin{equation}\label{T-C-det}
T+C=-\,\sqrt{2}\,\frac{V_{us}}{V_{ud}}\,
\frac{f_K}{f_{\pi}}\,A(B^+\to\pi^+\pi^0).
\end{equation}
Here the ratio $f_K/f_{\pi}=1.2$ of the kaon and pion decay constants
takes into account factorizable $SU(3)$-breaking corrections. Electroweak
penguin corrections to this expression can be taken into account 
theoretically \cite{BF,GPY}, but play a minor role in this case. The 
CLEO collaboration sees already some indication for $B^\pm\to\pi^\pm\pi^0$ 
modes \cite{CLEO1}, with a CP-averaged branching ratio~of 
\begin{equation}\label{Bpipi-res}
\mbox{BR}(B^\pm\to\pi^\pm\pi^0)=\left(5.6^{+2.6}_{-2.3}\pm1.7\right)\times
10^{-6}.
\end{equation}
However, the statistical significance of the signal yield is not yet 
sufficient to claim an observation of this channel. Using nevertheless
(\ref{Bpipi-res}), and taking into account the measured CP-averaged 
$B_d\to\pi^0K$ branching ratio, the combination of (\ref{rn-def}) and 
(\ref{T-C-det}) yields
\begin{equation}
r_{\rm n}=0.17\pm0.06\,,
\end{equation}
where we have added the experimental errors in quadrature.

The bounds on $\gamma$ implied by $R_{\rm n}$ are related to extremal
values of this observable. If we keep $r_{\rm n}$ and $\delta_{\rm n}$
as free parameters, we obtain the following minimal value for $R_{\rm n}$
\cite{defan}:
\begin{equation}\label{Rmin}
\left.R_{\rm n}^{\rm min}\right|_{r_{\rm n},\delta_{\rm n}}=
\left[\frac{1+2\,q\,\rho_{\rm n}\,\cos(\theta_{\rm n}+\omega)
+q^2\rho_{\rm n}^2}{\left(1-2\,q\,\cos\omega\cos\gamma+q^2\right)
\left(1+2\,\rho_{\rm n}\,\cos\theta_{\rm n}\cos\gamma+\rho_{\rm n}^2\right)}
\right]\sin^2\gamma\,.
\end{equation}
On the other hand, if only the strong phase $\delta_{\rm n}$ is kept as
an unknown quantity, $R_{\rm n}$ takes minimal and maximal values, which
are given by \cite{BF}
\begin{equation}\label{Rext}
\left.R_{\rm n}^{\rm ext}
\right|_{\delta_{\rm n}}=1\,\pm\,2\,\frac{r_{\rm n}}{u_{\rm n}}\,
\sqrt{h_{\rm n}^2+k_{\rm n}^2}\,+\,v^2r^2_{\rm n}.
\end{equation}
Expressions (\ref{Rmin}) and (\ref{Rext}) are the main 
equations of our paper. The parameter $\rho_{\rm n}$ is usually expected 
at the level of a few percent \cite{GHLR}, and governs also direct CP 
violation in $B_d\to\pi^0K$; model calculations of the corresponding CP 
asymmetry give results within the range $[0.4\%,5\%]$ \cite{AKL}. However, 
it should be kept in mind that $\rho_{\rm n}$ may be enhanced by 
final-state-interaction processes \cite{FSI}. These issues will be
discussed in more detail in the following section. 

The formulae given above apply also to the charged $B\to\pi K$ system,
if we perform the following replacements:
\begin{equation}\label{replace}
r_{\rm n}\to r_{\rm c}\equiv
\frac{|T+C|}{\sqrt{\langle\left|P\right|^2\rangle}},\quad
\rho_{\rm n}\,e^{i\theta_{\rm n}}\to \rho\,e^{i\theta},\quad \delta_{\rm n}
\to \delta_{\rm c}\equiv\delta_{T+C}-\delta_{tc}^{\rm c},
\end{equation}
where
\begin{equation}\label{Bpampl}
P\equiv A(B^+\to\pi^+K^0)=-\left(1-\frac{\lambda^2}{2}\right)\lambda^2A
\left[1+\rho\,e^{i\theta}e^{i\gamma}\right]\left|{\cal P}_{tc}^{\rm c}\right|
e^{i\delta_{tc}^{\rm c}}\,,
\end{equation}
with
\begin{equation}\label{rho-def}
\rho\,e^{i\theta}=\frac{\lambda^2R_b}{1-\lambda^2}
\left[1-\left(\frac{{\cal P}_{uc}^{\rm c}+
{\cal A}}{{\cal P}_{tc}^{\rm c}}\right)\right].
\end{equation}
Here the amplitude ${\cal A}$ is due to annihilation topologies. Using
(\ref{T-C-det}), (\ref{Bpipi-res}) and the measured CP-averaged
$B^\pm\to\pi^\pm K$ branching ratio, we obtain
\begin{equation}\label{rc-range}
r_{\rm c}=0.21\pm0.06,
\end{equation}
where we have again added the experimental errors in quadrature. 

The parameter $\rho$ is a measure of the importance of certain rescattering 
effects \cite{FSI}--\cite{FKNP}, and can be probed by comparing 
$B^\pm\to\pi^\pm K$ with its $U$-spin counterpart $B^\pm\to K^\pm K$ 
\cite{BFM,FKNP,defan}. To this end, we consider the following quantity
\begin{equation}\label{K-def}
K\equiv\left[\frac{1}{\epsilon\,R_{SU(3)}^2}\right]\left[
\frac{\mbox{BR}(B^\pm\to\pi^\pm K)}{\mbox{BR}(B^\pm\to K^\pm K)}\right]=
\frac{1+2\,\rho\cos\theta\cos\gamma+\rho^2}{\epsilon^2-2\,\epsilon\,\rho
\cos\theta\cos\gamma+\rho^2},
\end{equation}
where $\epsilon\equiv\lambda^2/(1-\lambda^2)$, and 
\begin{equation}
R_{SU(3)}=\frac{F_{B\pi}(M_K^2;0^+)}{F_{BK}(M_K^2;0^+)}
\end{equation}
describes factorizable $U$-spin-breaking corrections. If we use the model of 
Bauer, Stech and Wirbel \cite{BSW} to estimate the relevant form factors, 
we obtain $R_{SU(3)}={\cal O}(0.7)$. The expression on the right-hand side
of (\ref{K-def}) implies the following allowed range for $\rho$ (for a
detailed discussion, see \cite{RF-Bpipi} and \cite{pirjol}):
\begin{equation}\label{rho-range}
\frac{1-\epsilon\,\sqrt{K}}{1+\sqrt{K}}\leq\rho\leq
\frac{1+\epsilon\,\sqrt{K}}{|1-\sqrt{K}|}.
\end{equation}
The present CLEO data give $\mbox{BR}(B^\pm\to K^\pm K)/
\mbox{BR}(B^\pm\to\pi^\pm K)<0.3$ at $90\%$ C.L.\ \cite{CLEO1}. Consequently,
using (\ref{rho-range}), this upper bound implies $\rho<0.15$ for
$R_{SU(3)}=0.7$, and is not in favour of dramatic rescattering effects,
although the upper bound is still one order of magnitude above the usual 
model calculations, making use of arguments based on factorization. 

Let us finally note that the formalism discussed in this section can also 
be applied to the ``mixed'' $B_d\to\pi^\mp K^\pm$, $B^\pm\to\pi^\pm K$ system.
To this end, we have just to make appropriate replacements of variables,
involving certain amplitudes $T$ and $P_{\rm ew}^{\rm C}$, which measure 
colour-allowed tree-diagram-like and colour-suppressed electroweak penguin 
topologies, respectively. In order to fix $T$, arguments based on the 
factorization hypothesis have to be employed, and usually it is assumed that 
the colour-suppressed electroweak penguin amplitude $P_{\rm ew}^{\rm C}$ 
plays a very minor role. However, in contrast to (\ref{Ampl-def}), 
these quantities may be affected by rescattering processes.
An interesting approach, making use of a heavy-quark expansion for 
non-leptonic $B$ decays, was recently proposed in Ref.~\cite{BBNS}, which 
could help to reduce the uncertainties related to $T$ and $P_{\rm ew}^{\rm C}$.
It should also be useful to reduce the theoretical uncertainties of
$r_{\rm n}$, $r_{\rm c}$ and $q\,e^{i\omega}$, which are due to 
non-factorizable $SU(3)$-breaking corrections. Moreover, this approach allows 
also a calculation of the parameters $\rho_{\rm n}\,e^{i\theta_{\rm n}}$ and 
$\rho\,e^{i\theta}$. We will not consider the $B_d\to\pi^\mp K^\pm$, 
$B^\pm\to\pi^\pm K$ system further in this paper, and refer the reader to 
Refs.~\cite{BF,defan}, where detailed discussions can be found. Recently,
also the utility of $B_s\to\pi K$ decays in this context was pointed out 
\cite{BspiK}.

\begin{figure}
\centerline{\rotate[r]{
\epsfysize=11.2truecm
{\epsffile{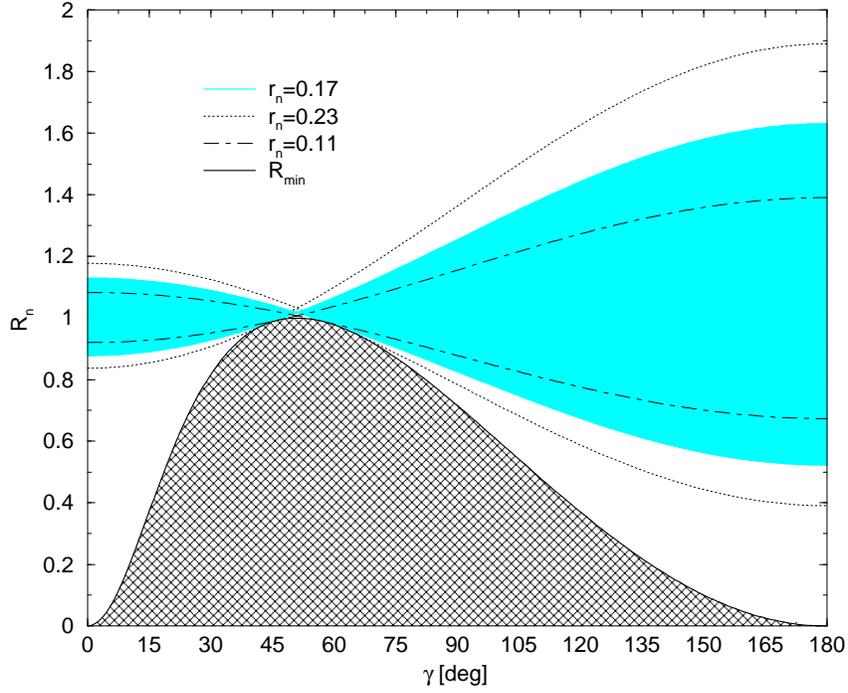}}}}
\caption{The dependence of the extremal values of $R_{\rm n}$ (neutral
$B\to\pi K$ system) described by
(\ref{Rmin}) and (\ref{Rext}) on the CKM angle $\gamma$ for 
$q e^{i\omega}=0.63$ and $\rho_{\rm n}=0$.}\label{fig:Rn}
\end{figure}

\begin{figure}
\centerline{\rotate[r]{
\epsfysize=11.2truecm
{\epsffile{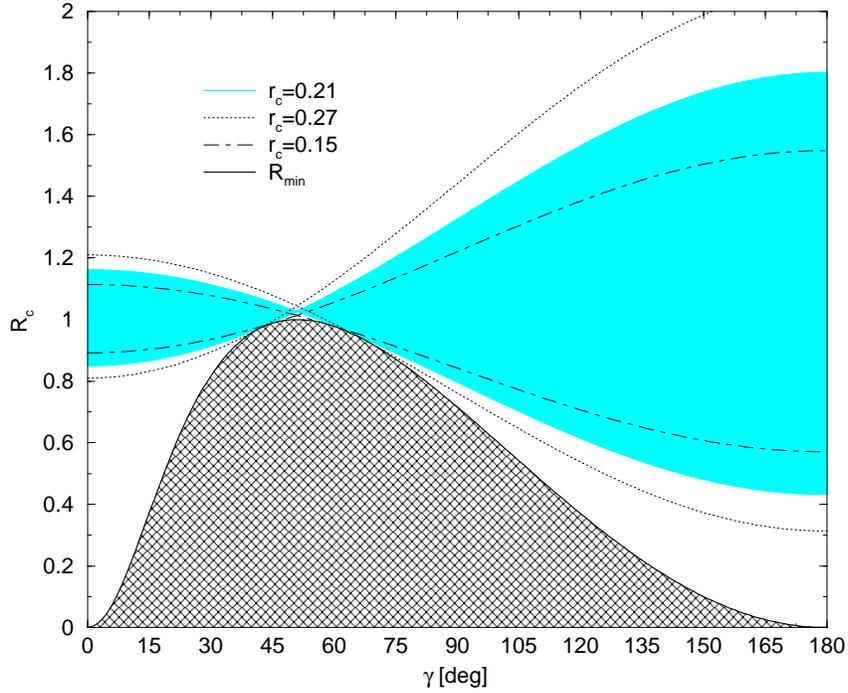}}}}
\caption{The dependence of the extremal values of $R_{\rm c}$ (charged
$B\to\pi K$ system) corresponding to (\ref{Rmin}) and (\ref{Rext}) on the 
CKM angle $\gamma$ for $q e^{i\omega}=0.63$ and $\rho=0$.}\label{fig:Rc}
\end{figure}

\boldmath
\section{Bounds on $\gamma$ and Constraints in the 
$\overline{\varrho}$--$\overline{\eta}$ Plane}\label{sec:bounds}
\unboldmath
The bounds on the CKM angle $\gamma$ implied by the CP-averaged branching
ratios of the neutral $B\to\pi K$ decays are related to the extremal
values of $R_{\rm n}$ given in (\ref{Rmin}) and (\ref{Rext}). In 
Fig.~\ref{fig:Rn}, we show their dependence on $\gamma$ for 
$q e^{i\omega}=0.63$ and $\rho_{\rm n}=0$.\footnote{In Fig.~\ref{fig:Rn},
we have assumed $0^\circ\leq\gamma\leq180^\circ$, as implied by the measured 
CP-violating parameter $\varepsilon_K$ of the neutral kaon system.} 
Here all values of $R_{\rm n}$ below the $R_{\rm min}$ curve are excluded. 
If $r_{\rm n}$ is fixed,
for example to be equal to 0.17, all values of $R_{\rm n}$ outside 
the shaded region are excluded, which is enlarged (reduced) for larger 
(smaller) values of $r_{\rm n}$. Fig.~\ref{fig:Rn} allows us to
read off immediately the allowed range for $\gamma$ corresponding
to a given value of $R_{\rm n}$. Let us consider, for example, the
central value of (\ref{Rn-def}), $R_{\rm n}=0.6$. In this case, the 
$R_{\rm min}$ curve implies the allowed range 
$0^\circ\leq\gamma\leq21^\circ\,\lor\,100^\circ\leq\gamma
\leq180^\circ$. If we use additional information on the parameter 
$r_{\rm n}$, we may put even stronger constraints on $\gamma$. For 
$r_{\rm n}=0.17$, we obtain, for instance, the allowed range 
$138^\circ\leq\gamma\leq180^\circ$. 

In the case of the charged $B\to\pi K$ system, bounds on $\gamma$ can
be obtained in an analogous manner. The corresponding curves for the
extremal values of $R_{\rm c}$ are shown in Fig.~\ref{fig:Rc}. There
is some kind of complementarity between the neutral and charged 
$B\to\pi K$ systems, since the CLEO data favour $R_{\rm n}<1$ and
$R_{\rm c}>1$. Consequently, we have to fix $r_{\rm c}$ in order
to constrain $\gamma$ through the charged $B\to\pi K$ decays. For
the central values of (\ref{Rc-def}) and (\ref{rc-range}), $R_{\rm c}=1.3$
and $r_{\rm c}=0.21$, we obtain $87^\circ\leq\gamma\leq180^\circ$. 

The allowed ranges for $\gamma$ arising in the examples given above would 
be of particular phenomenological interest, as they would be complementary 
to the range of $\gamma$ arising from the usual indirect fits 
of the unitarity triangle \cite{UT-fits}. The most recent analysis 
\cite{AL-recent} gives
\begin{equation}
38^\circ\leq\gamma\leq81^\circ.
\end{equation}
In our examples of the bounds from the neutral $B\to\pi K$ system, there 
would be no overlap between these ranges, which could be interpreted as a 
manifestation of new physics \cite{new-phys,GNK}. In particular, the second
quadrant for $\gamma$ is favoured; other arguments for $\cos\gamma<0$
using $B\to PP$, $PV$ and $VV$ decays were recently given in \cite{HY}
(see also \cite{RF-Bpipi}). However, the present data do not yet 
allow us to draw any definite conclusions. Before we can speculate on 
physics beyond the Standard Model, it is of course crucial to explore 
hadronic uncertainties. For the formalism used in this paper, this was done 
in \cite{BF}; within a different framework, similar considerations 
were also made for the charged and ``mixed'' $B\to\pi K$ systems in 
\cite{neubert}. 

The theoretical accuracy of the bounds on $\gamma$ discussed in this 
section is limited both by non-factorizable $SU(3)$-breaking corrections 
and by rescattering processes. The former may affect the determination of
the parameters $qe^{i\omega}$ and $r_{\rm n,c}$, whereas the latter
may lead to sizeable values of $\rho_{\rm n}$ and $\rho$. In order
to control the non-factorizable $SU(3)$-breaking corrections, the 
``QCD factorization'' approach presented in \cite{BBNS} appears to be 
very promising.

In the case of the neutral strategy, the parameter 
$\rho_{\rm n}e^{i\theta_{\rm n}}$ can be probed -- and even taken into 
account in the bounds on $\gamma$ in an {\it exact} manner --  through
CP-violating effects. To this end, we consider the $B_d\to\pi^0K$ modes
and require that the kaon be observed as a $K_{\rm S}$. The resulting final 
state is then an eigenstate of the CP operator with eigenvalue $-1$, and 
we obtain the following time-dependent CP asymmetry \cite{BF}:
\begin{eqnarray}
\lefteqn{a_{\rm CP}(B_d(t)\to\pi^0K_{\rm S})\equiv
\frac{\mbox{BR}(B_d^0(t)\to\pi^0K_{\rm S})\,-\,
\mbox{BR}(\overline{B^0_d}(t)\to\pi^0K_{\rm S})}{\mbox{BR}(B_d^0(t)\to
\pi^0K_{\rm S})\,+
\,\mbox{BR}(\overline{B^0_d}(t)\to\pi^0K_{\rm S})}}\nonumber\\
~~~~&=&{\cal A}_{\rm CP}^{\rm dir}(B_d\to\pi^0K_{\rm S})
\,\cos(\Delta M_d\,t)\,+\,{\cal A}_{\rm CP}^{\rm mix}(B_d\to\pi^0K_{\rm S})\,
\sin(\Delta M_d\,t)\,,\label{CPASY}
\end{eqnarray}
where ${\cal A}_{\rm CP}^{\rm dir}(B_d\to\pi^0K_{\rm S})$ and 
${\cal A}_{\rm CP}^{\rm mix}(B_d\to\pi^0K_{\rm S})$ are due to
``direct'' and ``mixing-induced'' CP violation, respectively. Using 
(\ref{Bnampl}), these observables take the following form:
\begin{equation}\label{Adir}
{\cal A}_{\rm CP}^{\rm dir}(B_d\to\pi^0K_{\rm S})=
-\,\frac{2\,\rho_{\rm n}\,\sin\theta_{\rm n}\sin\gamma}{1+
2\,\rho_{\rm n}\,\cos\theta_{\rm n}\cos\gamma+\rho_{\rm n}^2}
\end{equation}
\begin{eqnarray}
\lefteqn{{\cal A}_{\rm CP}^{\rm mix}(B_d\to\pi^0K_{\rm S})=}\nonumber\\
&&-\left[\frac{\sin\left(\phi_{\rm M}^{(d)}+\phi_K\right)+2\,\rho_{\rm n}
\cos\theta_{\rm n}\sin\left(\phi_{\rm M}^{(d)}+\phi_K+\gamma\right)+
\rho_{\rm n}^2\sin\left(\phi_{\rm M}^{(d)}+\phi_K+2\,\gamma\right)}{1+
2\,\rho_{\rm n}\cos\theta_{\rm n}\cos\gamma+\rho_{\rm n}^2}
\right].\label{Amix}
\end{eqnarray}
The latter expression reduces to
\begin{equation}\label{CP-rel}
{\cal A}_{\rm CP}^{\rm mix}(B_d\to\pi^0K_{\rm S})=-\,
\sin\left(\phi_{\rm M}^{(d)}+\phi_K\right)=
{\cal A}_{\rm CP}^{\rm mix}(B_d\to J/\psi\,K_{\rm S})
\end{equation} 
in the case of $\rho_{\rm n}=0$ \cite{PAPIII}. Clearly, a violation of 
(\ref{CP-rel}) and a sizeable value of the direct CP asymmetry (\ref{Adir}) 
would signal that the parameter $\rho_{\rm n}$ cannot be neglected. Such
a feature may either be due to large rescattering effects, or to new-physics
contributions. The whole pattern of all experimentally observed $B\to\pi K$ 
and $B\to K\overline{K}$ decays may allow us to distinguish between these 
cases. 

In the mixing-induced CP asymmetry (\ref{CP-rel}), 
$\phi_{\rm M}^{(d)}=2\,\mbox{arg}(V_{td}^\ast V_{tb})$ is 
related to the weak $B^0_d$--$\overline{B^0_d}$ mixing phase, whereas 
$\phi_K$ is related to $K^0$--$\overline{K^0}$ mixing, and is negligibly 
small in the Standard Model. The combination 
$\phi_d=\phi_{\rm M}^{(d)}+\phi_K$ is equal to $2\beta$ in the Standard 
Model, and can be determined ``straightforwardly'' through the  
``gold-plated'' mode $B_d\to J/\psi\,K_{\rm S}$ at the $B$-factories. 
Strictly speaking, a measurement of 
${\cal A}_{\rm CP}^{\rm mix}(B_d\to J/\psi\,K_{\rm S})$ 
allows us to determine only $\sin\phi_d$, i.e.\ to fix $\phi_d$ up to a 
twofold ambiguity. Several strategies were proposed in the literature to 
resolve this ambiguity \cite{ambig}. 

\begin{figure}
\centerline{\rotate[r]{
\epsfysize=13.1truecm
{\epsffile{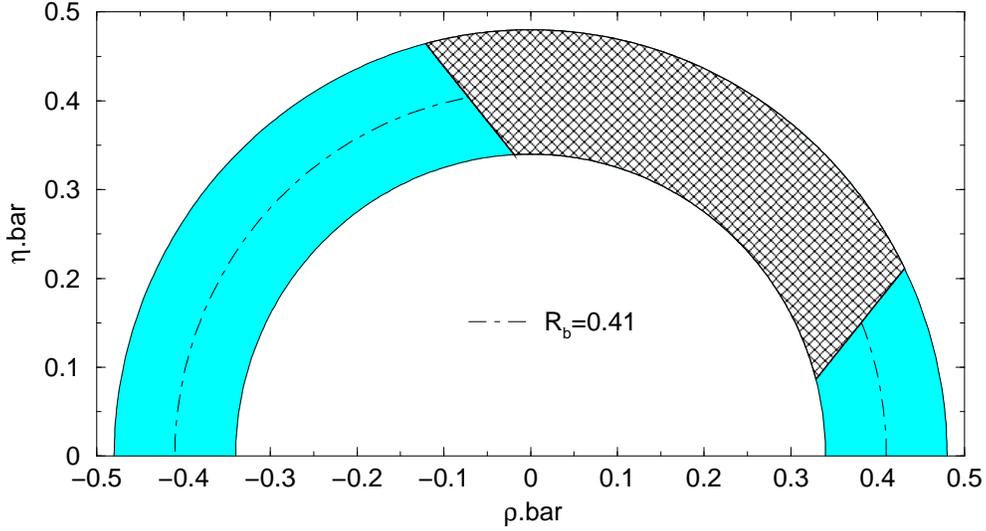}}}}
\caption{The constraints in the $\overline{\varrho}$--$\overline{\eta}$ 
plane implied by (\ref{Rmin}) for $R_{\rm n}=0.6$, $q e^{i\omega}=0.63
\times[0.41/R_b]$, and $\rho_{\rm n}=0$. The shaded region is the allowed 
range for the apex of the unitarity triangle, whereas the ``crossed'' region 
is excluded through $R_{\rm n}^{\rm min}|_{r_{\rm n},\delta_{\rm n}}$ 
(see Fig.~\ref{fig:Rn}).}\label{fig:rho-eta1}
\end{figure}

\begin{figure}
\centerline{\rotate[r]{
\epsfysize=13.1truecm
{\epsffile{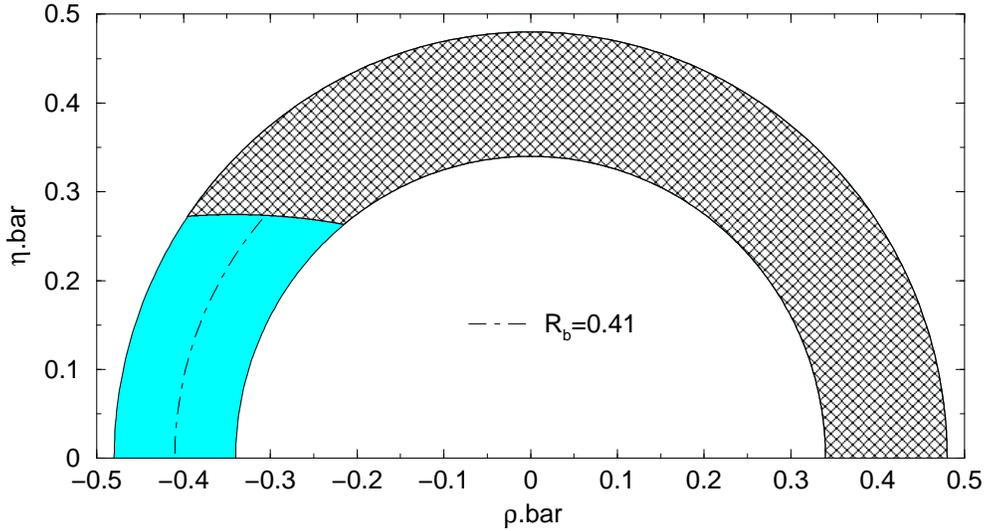}}}}
\caption{The constraints in the $\overline{\varrho}$--$\overline{\eta}$ 
plane implied by (\ref{Rext}) for $R_{\rm n}=0.6$, $r_{\rm n}=0.17$,
$q e^{i\omega}=0.63\times[0.41/R_b]$, and $\rho_{\rm n}=0$. The shaded 
region is the allowed range for the apex of the unitarity triangle, whereas 
the ``crossed'' region is excluded through 
$R_{\rm n}^{\rm ext}|_{\delta_{\rm n}}$ (see 
Fig.~\ref{fig:Rn}).}\label{fig:rho-eta2}
\end{figure}

If we assume that $\phi_d$ has been fixed this way, the observables 
(\ref{Adir}) and (\ref{Amix}) allow us to determine $\rho_{\rm n}$ and 
$\theta_{\rm n}$ as a function of $\gamma$. The general formulae given 
in the previous section allow us then to take into account these parameters 
in the curves shown in Fig.~\ref{fig:Rn}. The usual model calculations for 
non-leptonic $B$ decays give values for $\rho_{\rm n}$ at the level of a 
few percent. In order to illustrate the impact on the bounds on $\gamma$, 
let us take $\rho_{\rm n}=0.05$ and $\theta_{\rm n}\in\{0^\circ,180^\circ\}$. 
For the example given above, we obtain then the allowed ranges 
$0^\circ\leq\gamma\leq\left(21^\circ\pm1^\circ\right)\,\lor\,
\left(100^\circ\pm4^\circ\right)\leq\gamma\leq180^\circ$, and
$\left(138^\circ\pm2^\circ\right)\leq\gamma\leq180^\circ$.
The feature that the uncertainty due to $\rho_{\rm n}$ is larger in the 
case of $R^{\rm min}_{\rm n}$ can be understood easily by performing
an expansion of (\ref{Rmin}) and (\ref{Rext}) in powers of $\rho_{\rm n}$,
and neglecting second-order terms of ${\cal O}(\rho_{\rm n}^2)$,
${\cal O}(r_{\rm n}\,\rho_{\rm n})$ and ${\cal O}(r_{\rm n}^2)$:
\begin{eqnarray}
\left.R_{\rm n}^{\rm min}\right|_{r_{\rm n},\delta_{\rm n}}^{\rm L.O.}&=&
\left[\frac{1+2\,\rho_{\rm n}\cos\theta_{\rm n}\left(q-\cos\gamma
\right)}{1-2\,q\,\cos\gamma+q^2}\right]\sin^2\gamma\label{Rmin-approx}\\
\left.R_{\rm n}^{\rm ext}
\right|_{\delta_{\rm n}}^{\rm L.O.}&=&1\,\pm\,2\,r_{\rm n}
\left|\cos\gamma-q\right|.\label{Rext-approx}
\end{eqnarray}
Here we have moreover made use of (\ref{q-expr}), which gives
$\omega=0$. Interestingly, as was noted for the charged $B\to\pi K$ system 
in \cite{NR}, there are no terms of ${\cal O}(\rho_{\rm n})$ present in 
(\ref{Rext-approx}), in contrast to (\ref{Rmin-approx}). Consequently, 
the bounds on $\gamma$ related to (\ref{Rmin}) are affected more strongly 
by $\rho_{\rm n}$ then those implied by (\ref{Rext}). In the case of the 
charged strategy, we have to use the $U$-spin flavour symmery, relating 
$B^\pm\to\pi^\pm K$ to $B^\pm\to K^\pm K$, in order to take into account 
the parameters $\rho$ and $\theta$ in the curves shown in 
Fig.~\ref{fig:Rc} \cite{BF,defan}. To this end, the observable $K$ 
introduced in (\ref{K-def}) has to be combined with the direct CP 
asymmetries in $B^\pm\to\pi^\pm K$ or $B^\pm\to K^\pm K$ modes.

In addition to the theoretical uncertainties associated with $SU(3)$-breaking
and rescattering effects, another uncertainty of the constraints 
on $\gamma$ is due to the CKM factor $R_b$ in expression (\ref{q-expr}) 
for the electroweak penguin parameter $qe^{i\omega}$. Because of this feature,
it is actually more appropriate to consider constraints in the 
$\overline{\varrho}$--$\overline{\eta}$ plane. A similar ``trick'' was 
also employed for $B_d\to\pi^+\pi^-$ decays in \cite{charles}, and 
recently for the charged $B\to\pi K$ system in \cite{neubert-proc}.

The constraints in the $\overline{\varrho}$--$\overline{\eta}$
plane can be obtained straightforwardly from (\ref{Rmin}) and (\ref{Rext}).
In the former case, we obtain
\begin{equation}\label{constr1}
\cos\gamma=R_{\rm n}q\pm\sqrt{\left(1-R_{\rm n}\right)\left(
1-R_{\rm n}q^2\right)},
\end{equation}
whereas we have in the latter case
\begin{equation}\label{constr2}
\cos\gamma=\frac{1-R_{\rm n}\pm2\,q\,r_{\rm n}+\left(1+
q^2\right)r_{\rm n}^2}{2\,r_{\rm n}\left(q\,r_{\rm n}\pm1\right)}.
\end{equation}
In these expressions, we have assumed, for simpliciy, $\rho_{\rm n}=0$ and 
$\omega=0$. For the charged $B\to\pi K$ system, we obtain analogous 
expressions. The right-hand sides of these formulae depend implicitly on 
the CKM factor $R_b$ through the electroweak penguin parameter 
$qe^{i\omega}$, which is given by (\ref{q-expr}). Consequently, it is 
actually more appropriate to consider contours in the 
$\overline{\varrho}$--$\overline{\eta}$ plane instead of the CKM angle 
$\gamma$. They can be obtained with the help of (\ref{constr1}) and 
(\ref{constr2}) by taking into account \cite{BLO}
\begin{equation}
\overline{\varrho}=R_b\cos\gamma,\quad\overline{\eta}=R_b\sin\gamma,
\end{equation}
and are illustrated in Figs.~\ref{fig:rho-eta1} and \ref{fig:rho-eta2}
for the examples given in the previous section.

\begin{figure}
\centerline{\rotate[r]{
\epsfysize=11.2truecm
{\epsffile{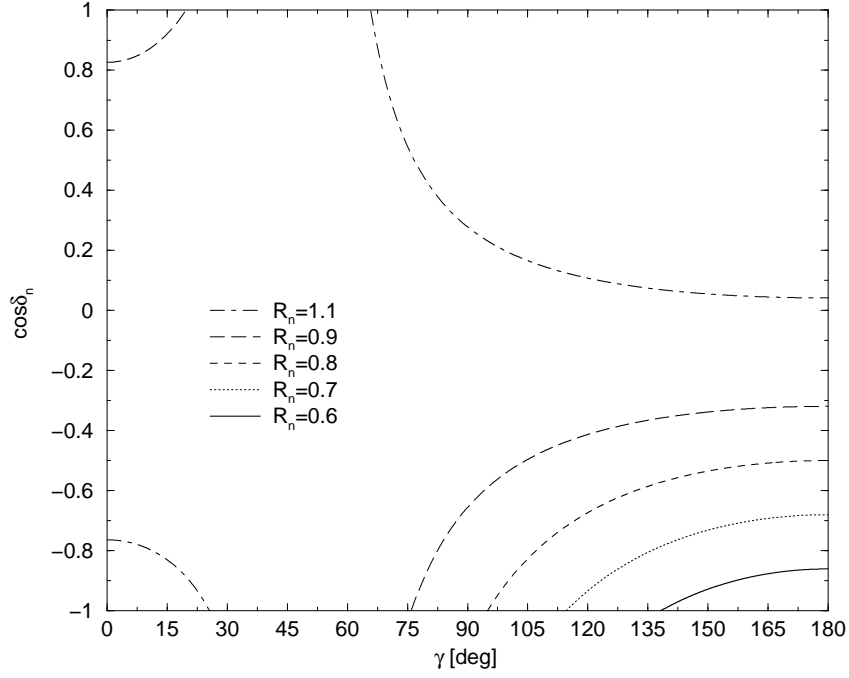}}}}
\caption{The dependence of $\cos\delta_{\rm n}$ on $\gamma$ for various 
values of $R_{\rm n}$ in the case of $q e^{i\omega}=0.63$ and 
$r_{\rm n}=0.17$. Rescattering effects are neglected, i.e.\ 
$\rho_{\rm n}=0$.}\label{fig:cos-n}
\end{figure}

\begin{figure}
\centerline{\rotate[r]{
\epsfysize=10.9truecm
{\epsffile{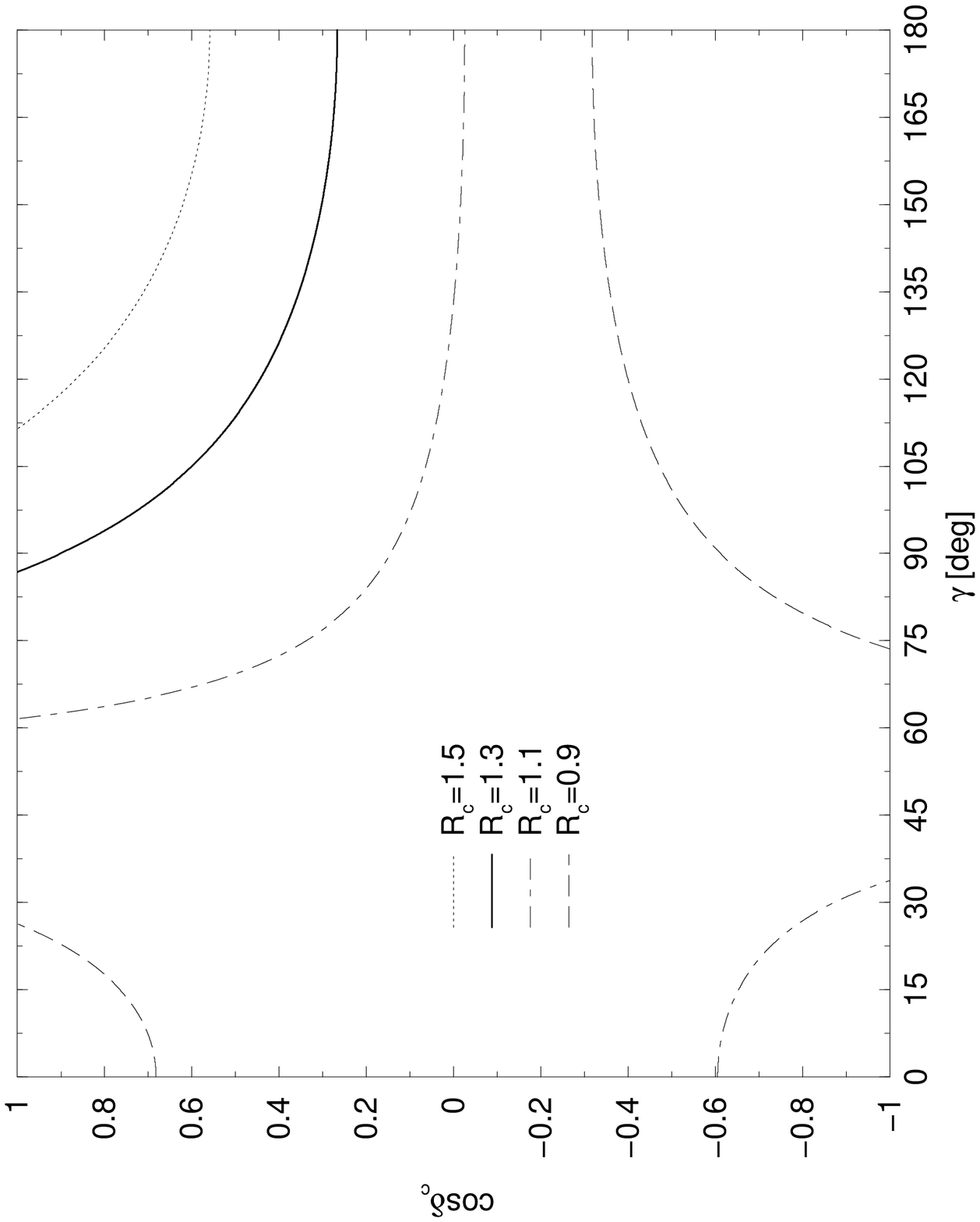}}}}
\caption{The dependence of $\cos\delta_{\rm c}$ on $\gamma$ for various 
values of $R_{\rm c}$ in the case of $q e^{i\omega}=0.63$ and 
$r_{\rm c}=0.21$. Rescattering effects are neglected, i.e.\ 
$\rho=0$.}\label{fig:cos-c}
\end{figure}

\begin{figure}
\centerline{\rotate[r]{
\epsfysize=10.9truecm
{\epsffile{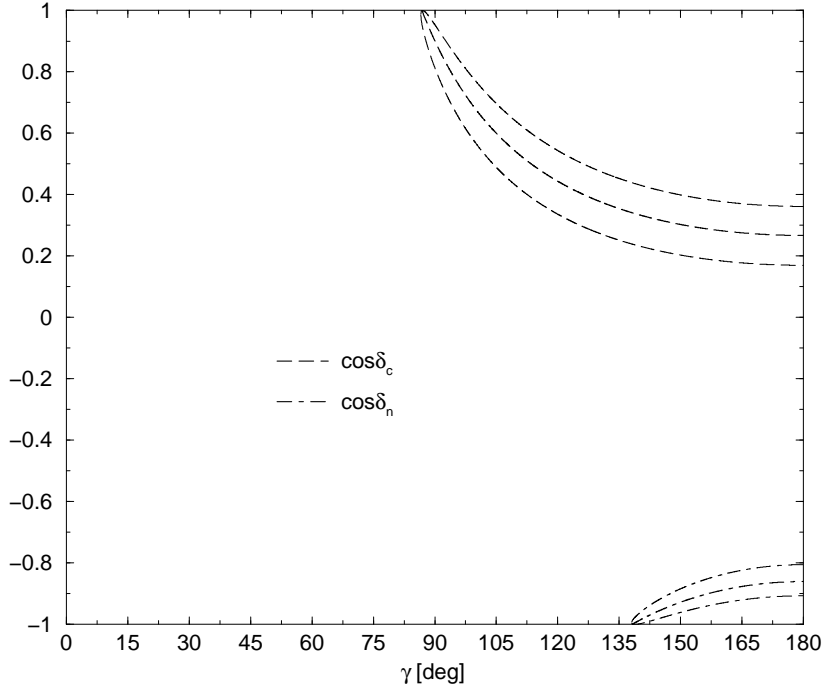}}}}
\caption{The dependence of $\cos\delta_{\rm c}$ and $\cos\delta_{\rm n}$
on $\gamma$ for $R_{\rm c}=1.3$, $r_{\rm c}=0.21$, $R_{\rm n}=0.6$, 
$r_{\rm n}=0.17$, $q e^{i\omega}=0.63$ in the presence of large rescattering
effects (thin lines), corresponding to $\rho e^{i\theta}=
\rho_{\rm n}e^{i\theta_{\rm n}}=
0.1\times\exp(i\, 90^\circ)$.}\label{fig:cos-FSI}
\end{figure}

\begin{figure}
\centerline{\rotate[r]{
\epsfysize=11.2truecm
{\epsffile{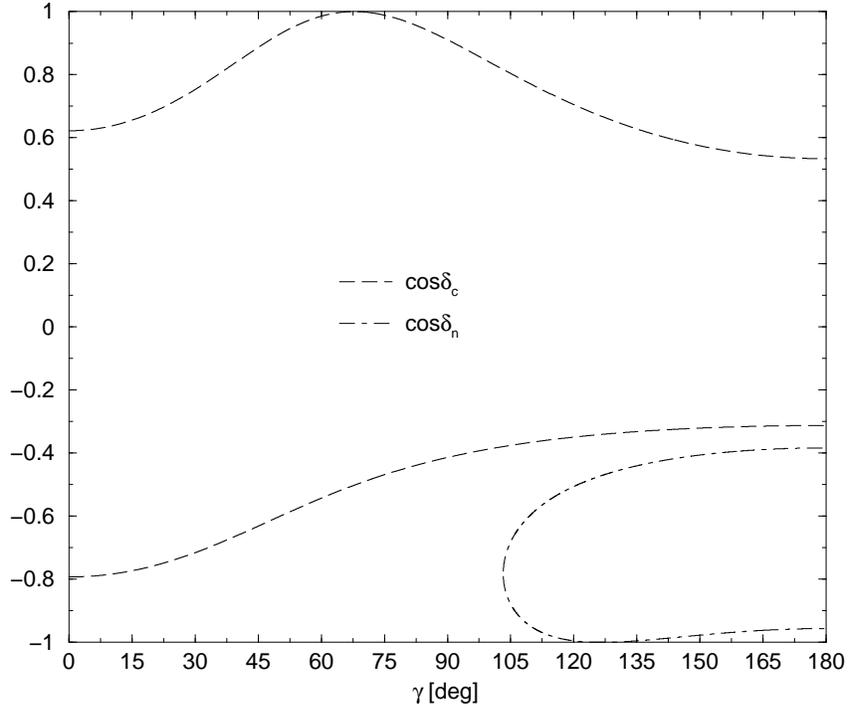}}}}
\caption{The dependence of $\cos\delta_{\rm c}$ and $\cos\delta_{\rm n}$
on $\gamma$ for $R_{\rm c}=1.3$, $r_{\rm c}=0.21$, $R_{\rm n}=0.6$, 
$r_{\rm n}=0.17$, $\rho=\rho_{\rm n}=0$ for a modified electroweak penguin 
parameter, given by $q e^{i\omega}=1.26\times
\exp(i\, 45^\circ)$.}\label{fig:cos-ewp}
\end{figure}

\boldmath
\section{Bounds on Strong Phases}\label{sec:delta}
\unboldmath
If we use the general expression (\ref{Rn-exp}) for $R_{\rm n}$, we
can determine $\cos\delta_{\rm n}$ as a function of $\gamma$:
\begin{equation}\label{cosdelta}
\cos\delta_{\rm n}=\frac{1}{h_{\rm n}^2+k_{\rm n}^2}\left[
\frac{\left(1-R_{\rm n}+v^2r_{\rm n}^2\right)u_{\rm n}h_{\rm n}}{2\,r_{\rm n}}
\pm k_{\rm n}\sqrt{h_{\rm n}^2+k_{\rm n}^2-\left[\frac{\left(1-R_{\rm n}+
v^2r_{\rm n}^2\right)u_{\rm n}}{2\,r_{\rm n}}\right]^2}\right].
\end{equation}
In Fig.~\ref{fig:cos-n}, we show the dependence of $\cos\delta_{\rm n}$ 
for various values of $R_{\rm n}$ in the case of $q e^{i\omega}=0.63$ and 
$r_{\rm n}=0.17$. From this figure, also the allowed range for $\gamma$ can
be read off for a given value of $R_{\rm n}$. For the central value 
$R_{\rm n}=0.6$ of the present CLEO data, we obtain moreover $-1\leq
\cos\delta_{\rm n}\leq-0.86$. Performing the replacements given in
(\ref{replace}), (\ref{cosdelta}) applies also to the charged $B\to\pi K$ 
system. The corresponding contours in the $\gamma$--$\cos\delta_{\rm c}$ 
plane are shown in Fig.~\ref{fig:cos-c}. For $R_{\rm c}=1.3$, we obtain
$+0.27\leq\cos\delta_{\rm c}\leq+1$. 

As can be seen in (\ref{rn-def}) and (\ref{replace}), we have 
$\delta_{\rm n}-\delta_{\rm c}=\delta_{tc}^{\rm c}-\delta_{tc}^{\rm n}$, 
where $\delta_{tc}^{\rm c}$ and $\delta_{tc}^{\rm n}$ denote the strong 
phases of the amplitudes ${\cal P}_{tc}^{\rm c}$ and ${\cal P}_{tc}^{\rm n}$,
which describe the differences of penguin topologies with internal top- 
and charm-quark exchanges of the decays $B^+\to\pi^+K^0$ and 
$B_d^0\to\pi^0K^0$, respectively. These penguin topologies consist of QCD 
and electroweak penguins, where the latter contribute to $B^+\to\pi^+K^0$ 
only in colour-suppressed form. In contrast, $B_d^0\to\pi^0K^0$ receives 
contributions both from colour-allowed and from colour-suppressed electroweak 
penguins. Nevertheless, they are expected to be at most of 
${\cal O}(20\%)$ of the $B_d^0\to\pi^0K^0$ QCD penguin amplitude. If we 
neglect the electroweak penguins and make use of isospin flavour-symmetry 
arguments, we obtain ${\cal P}_{tc}^{\rm n}\approx{\cal P}_{tc}^{\rm c}$, 
yielding $\delta_{\rm n}\approx\delta_{\rm c}$ and $\cos\delta_{\rm n}\approx
\cos\delta_{\rm c}$. Employing moreover ``factorization'', these cosines
are expected to be close to $+1$. 

Consequently, as the present CLEO data are in favour of $\cos\delta_{\rm n}<0$
and $\cos\delta_{\rm c}>0$, we arrive at a ``puzzling'' situation, although it
is of course too early to draw definite conclusions. If future data should 
confirm this ``discrepancy'', it may be an indication for new-physics 
contributions to the electroweak penguin sector, or a manifestation of 
large non-factorizable $SU(3)$-breaking effects. Since the parameter 
$\rho_{\rm n}$ enters in expression (\ref{Rn-exp}) for $R_{\rm n}$ in the 
term proportional to $r_{\rm n}$, it can be regarded as a second-order 
effect and does not play a dramatic role for the contraints on 
$\cos\delta_{\rm n(c)}$ and $\gamma$. This feature is illustrated in 
Fig.~\ref{fig:cos-FSI} for the central values of the present CLEO data. 

In Fig.~\ref{fig:cos-ewp}, we consider the impact of a modified electroweak
penguin parameter, $q e^{i\omega}=1.26\times\exp(i\, 45^\circ)$, which
differs significantly from the $SU(3)$ Standard-Model expression 
(\ref{q-expr}). In this case, the discrepancy between $\cos\delta_{\rm n}$ 
and $\cos\delta_{\rm c}$ would be essentially resolved, favouring values 
of ${\cal O}(-0.5)$, which would still be in conflict with the factorization 
expectation. A value of $q e^{i\omega}=1.26\times\exp(i\, 45^\circ)$ may be 
due to CP-conserving new-physics contributions to the electroweak penguin
sector \cite{GNK}. In general, new physics will also lead to CP-violating 
contributions, which may lead to sizeable direct CP violation in 
$B_d\to\pi^0K_{\rm S}$, and to a violation of (\ref{CP-rel}). Consequently, 
as we have already emphasized above, it would be an important task to 
measure the CP-violating observables of this decay. 

If the new-physics contributions are CP-conserving, it will be hard to 
distinguish them from large non-factorizable flavour-symmetry-breaking 
effects, which may also shift the parameter $q e^{i\omega}$ from 
(\ref{q-expr}). In Refs.\ \cite{GNK,neubert}, it was argued that these 
effects are very small, whereas we gave a more critical picture in \cite{BF}. 
Also the approach proposed in \cite{BBNS} is in favour of small 
non-factorizable effects. The deviation of 
$q e^{i\omega}=1.26\times\exp(i\, 45^\circ)$ used in the 
example given in Fig.~\ref{fig:cos-ewp} from (\ref{q-expr}) would probably 
be too large to be explained by $SU(3)$ breaking in a ``natural'' way. 
However, there may be additional sources for flavour-symmetry-breaking 
effects. An example is $\pi^0$--$\eta$,\,$\eta'$ mixing, which has not yet 
been considered for $B\to\pi^0 K$ decays. In a recent paper \cite{gardner}, 
it was emphasized that isospin violation arising from such effects could mock 
new physics in the extraction of the CKM angle $\alpha$ from $B\to\pi\pi$ 
isospin relations. It would be interesting to extend these studies also to 
the $B\to\pi K$ approaches to probe $\gamma$.

\section{Conclusions}\label{sec:concl}
As we have pointed out in Ref.~\cite{BF}, the neutral $B\to\pi K$ strategy 
could be useful to constrain -- and eventually determine -- $\gamma$ 
in an analogous manner as the strategy of Neubert and Rosner~\cite{NR}
using charged $B\to\pi K$ modes. The most recent CLEO data look very 
interesting in this respect. As we have illustrated in 
Figs.~\ref{fig:Rn}--\ref{fig:rho-eta2}, improved measurements of both the 
neutral and the charged modes, in particular taken together, could give a 
powerful constraint on $\gamma$. There is some indication that the second 
quadrant for $\gamma$ is preferred. This is in contrast to the standard 
analysis of the unitarity triangle, which favours the first quadrant. 
Unfortunately, no definite conclusions can be drawn at present. 
This ``discrepancy'' between the $B\to\pi K$ approaches and the standard 
analysis of the unitarity triangle could turn out to be more pronounced 
when the $B$-decay data improve and the lower bound on 
$B_s^0$--$\overline{B_s^0}$ mixing will be raised, forcing the upper bound 
on $\gamma$ from the standard analysis to be even smaller than presently 
known. 
   
We have also pointed out that the CLEO data suggest bounds on the strong 
phases $\delta_{\rm n}$ and $\delta_{\rm c}$ with $\cos\delta_{\rm n}<0$ 
and $\cos\delta_{\rm c}>0$. The substantial deviation of $\delta_{\rm n}$ 
from $\delta_{\rm c}$ and the negative value of $\cos\delta_{\rm n}$, if 
confirmed by improved data, would either indicate substantial new-physics 
contributions to the electroweak penguin sector, or large non-factorizable
$SU(3)$-breaking effects. In order to distinguish between these possibilties, 
detailed studies of the various patterns of new-physics effects in all 
$B\to\pi K$ decays are essential, as well as critical analyses of possible 
sources for $SU(3)$ breaking. We hope that future studies following the 
strategies discussed in this paper will eventually shed light on the 
physics beyond the Standard Model.

\vspace*{0.3truecm}

\noindent
This work has been supported in part by the German Bundesministerium f\"ur 
Bildung und Forschung under contract 05HT9WOA0.

\end{document}